\documentclass[aps,prl,twocolumn,showpacs,superscriptaddress,groupedaddress]{revtex4}  
\usepackage{graphicx}  
\usepackage{dcolumn}   
\usepackage{bbm}        
\usepackage{amssymb}   
\usepackage{amsmath}
\usepackage{enumerate}

\newcommand\seq[3]{{#1}_{#2}^{#3}}
\newcommand{\bd}{\begin{document}}
\newcommand{\ed}{\end{document}}
\newcommand{\beq}{\begin{equation}}
\newcommand{\eeq}{\end{equation}}
\newcommand{\bef}{\begin{figure}}
\newcommand{\enf}{\end{figure}}
\newcommand{\bea}{\begin{eqnarray}}
\newcommand{\eea}{\end{eqnarray}}
\newcommand{\baR}{\begin{array}}
\newcommand{\eaR}{\end{array}}
\newcommand{\bc}{\begin{center}}
\newcommand{\ec}{\end{center}}
\newcommand{\ben}{\begin{enumerate}}
\newcommand{\een}{\end{enumerate}}
\newcommand{\bit}{\begin{itemize}}
\newcommand{\eit}{\end{itemize}}
\newcommand{\su}{\section}
\newcommand{\ssu}{\subsection}
\newcommand{\sssu}{\subsubsection}
\newcommand\ssssu[1]{\textbf{#1}}
\newcommand{\nid}{\noindent}
\newcommand{\nnb}{\nonumber}
\newcommand\bloc[2]{{\omega}_{#1}^{#2}}
\newcommand\blocp[2]{{\omega'}_{#1}^{#2}}
\newcommand{\un}{\mathbb{1}}
\newcommand{\setZ}{\mathbbm{Z}}
\newcommand{\setN}{\mathbbm{N}}
\newcommand{\setR}{\mathbbm{R}}
\newcommand{\pare}[1]{\left(\, #1 \, \right)}
\newcommand{\scal}[2]{\langle\, #1 | #2 \, \rangle}
\newcommand{\bra}[1]{\left[\, #1 \, \right]}
\newcommand{\brac}[2]{\left[\, #1 \, | \, #2 \right]}
\newcommand{\Set}[1]{\left\{\, #1 \, \right\}}
\newcommand{\Setc}[2]{\left\{\, #1 \, ; \, #2 \right\}}
\newcommand{\Prob}[1]{P\left[\, #1 \, \right]}
\newcommand{\Probt}[2]{P_{#1}\left[\, #2 \, \right]}
\newcommand{\PT}[1]{\pi^{(T)}\left[\, #1 \, \right]}
\newcommand{\PTo}[1]{\pi^{(T)}_\omega\left[\, #1 \, \right]}
\newcommand{\pT}{\pi^{(T)}}
\newcommand{\Probex}[1]{P^{(T)}\left[\, #1 \, \right]}
\newcommand{\probc}[2]{P[#1 \, \left| \, #2 \right.]}
\newcommand{\Probc}[2]{P\bra{#1 \, \left| \, #2 \right.}}
\newcommand{\Probct}[3]{P_{#1}\bra{#2 \, \left| \, #3 \right.}}
\newcommand{\Probcex}[2]{P^{(T)}\bra{#1 \, \left| \, #2 \right.}}
\newcommand{\Probcp}[2]{P^{(ap)}\bra{#1 \, \left| \, #2 \right.}}
\newcommand{\abs}[1]{\left|\, #1 \, \right|}
\newcommand{\deq}{\stackrel {\rm def}{=}}
\newcommand{\sqsubsetneq}{\stackrel {\tiny\sqsubset}{\tiny \neq}}
\newcommand\moy[1]{\mu\bra{#1}}
\newcommand\noy[1]{\nu\bra{#1}}
\newcommand\cB{{\mathcal{B}}}
\newcommand\cC{{\mathcal{C}}}
\newcommand\cD{{\mathcal{D}}}
\newcommand\cT{{\mathcal{T}}}
\renewcommand\H{{\cH}}
\newcommand{\h}{\mathtt{h}}
\newcommand{\J}{\mathtt{J}}
\newcommand\s[1]{{\cS\bra{#1}}}
\newcommand\tH{{\tilde{H}}}
\renewcommand\th{{\tilde{h}}}
\newcommand\tFi{{\tilde{\Phi}}}
\newcommand\tfi{{\tilde{\phi}}}
\newcommand\tF{{\tilde{F}}}
\newcommand\tG{{\tilde{G}}}
\newcommand\tu{{\tilde{u}}}
\newcommand\tv{{\tilde{v}}}
\renewcommand{\sb}{s_\bbeta}
\newcommand\cE{{\mathcal{E}}}
\newcommand\cG{{\cal G}}
\newcommand\cH{{\mathcal{H}}}
\newcommand\cI{{\mathcal{I}}}
\newcommand\cK{{\cal K}}
\newcommand\cL{{\cal L}}
\newcommand\cM{{\mathcal{M}}}
\newcommand\cS{{\cal S}}
\newcommand\cP{{\mathcal{P}}}
\newcommand\cO{{\cal O}}
\newcommand\cU{{\cal U}}
\newcommand\cW{{\cal W}}
\newcommand\f{{\bm{f}}}
\newcommand\be{{\bm{e}}}
\newcommand\bh{{\bm{h}}}
\newcommand\bxi{{\bm{\xi}}}
\newcommand\bn{{\bm{n}}}
\newcommand\bv{{\bm{v}}}
\newcommand\bF{{\bm{F}}}
\newcommand\bo{{\bm{o}}}
\newcommand\blambda{{\bm{\lambda}}}
\newcommand\bbeta{{\bm{\beta}}}
\newcommand\gk[1]{g_k\pare{#1}}
\newcommand\gLk{g_{L,k}}
\newcommand\akj[1]{\alpha_{kj}(#1)}
\newcommand\sif[1]{\seq{\omega}{#1}{D}}
\newcommand\Xk[1]{X_k({#1})}
\newcommand\Ik[1]{I_k({#1})}
\newcommand\Skj[1]{S_{kj}({#1})}
\newcommand\X[2]{X_{#1}\pare{#2}}
\newcommand\et[2]{\eta_{#1}\pare{#2}}
\newcommand\tO[2]{\tau_{#1}\pare{#2}}
\newcommand\Vd[2]{\mathcal{V}^{(det)}_{#1}\pare{#2}}
\newcommand\XkD{\Xk{\bloc{0}{D-1}}}
\newcommand\Xkr{X_k^r}
\newcommand\tLk{\tau_{L,k}}
\newcommand\moyc[2]{\mu\bra{#1 \, | \, #2}}
\newcommand\Vkdet[1]{V_k^{(det)}({#1})}
\newcommand\Vksyn[1]{V_k^{(syn)}({#1})}
\newcommand\Vkext[1]{V_k^{(ext)}({#1})}
\newcommand\Vknoise[1]{V_k^{(noise)}({#1})}
\newcommand\sk[1]{\sigma_k({#1})}
\newcommand\skr{\sigma_k^r\pare{\bloc{0}{D-1}}}
\newcommand{\ent}[1]{\left[\, #1 \, \right]}
\newcommand\vm[1]{var_m\left[#1 \right]}
\newcommand\eg[1]{\stackrel {\rm #1}{=}}
\newcommand\tk[1]{\tau_k\pare{#1}}
\newcommand\tko{\tau_k(t,\omega)}
\newcommand\cBm[1]{\cB^{-1}\pare{#1}}
\newtheorem{theorem}{Theorem}
\newcommand{\bth}{\begin{theorem}}
\newcommand{\enth}{\end{theorem}}
\newcommand\omD[1]{\seq{\bra{\omega^{(#1)}}}{0}{D-1}}
\newcommand\omz[1]{\omega^{(#1)}(D)}
\newcommand\om[1]{\omega^{(#1)}}
\newcommand\Om[1]{\Omega^{(#1)}}
\newcommand\skjq{s_{kj;q}}
\newcommand\ikq{i_{k;q}}
\newcommand\skjqs{p_{kj;q_s}}
\newcommand\ikqs{i_{k;q_s}}
\newcommand\ikqsu{i_{k;q_{s_u}}}
\newcommand\ikqu{i_{k;q_u}}
\newcommand\xkq{x_{k;q}}
\newcommand\skjqv{s_{kj;q_{v}}}
\newcommand\tjr{t_j^{(r)}(\omega)}
\renewcommand\S[2]{S_{#1}\pare{#2}}
\newcommand{\im}{Im}
\newcommand\rpf{R}
\newcommand\rpfc[1]{R\pare{#1}}
\newcommand\lpf{L}
\newcommand\lpfc[1]{L\pare{#1}}
\newcommand{\zb}{\zeta_{\bbeta}}
\newcommand{\etab}{\eta_{\bbeta}}
\newcommand{\mex}{\mu^{\ast}}
\newcommand{\phex}{\phi^{\ast}}
\newcommand{\Phex}{\Phi^{\ast}}
\newcommand{\Hex}{\cH^{\ast}}
\newcommand{\hex}{h^{\ast}}
\newcommand{\map}{\mu^{(\chi)}}
\newcommand{\Hap}{\cH^{(\chi)}}
\newcommand{\hap}{h^{(\chi)}}
\newcommand{\siex}{\sigma^{\ast}}
\newcommand{\cn}{\cC^{\ast}}
\newcommand\p[1]{\cP\bra{#1}}
\newcommand\w[2]{w_{#1}^{(#2)}}
\newcommand\Gb{\cG_\bbeta}
\newcommand\K[2]{\cK_{\small{#1; \,  #2}}}
\renewcommand\d[2]{\delta_{\small{#1; \,  #2}}}
\newcommand\blp[3]{\omega^{#3,(#1)}_{#2}}
\newcommand\blpb[3]{\overline{\omega^{#3,(#1)}_{#2}}}
\newcommand\moyex[1]{\mu^{\ast}\bra{#1}}
\newcommand\E[1]{{\cal E}_{#1}}
\newcommand{\tc}{\tau(\cC)}
\newcommand{\notsqsupset}{\cancel{\sqsupset}}
\newcommand{\sm}{\sigma^{-1}}
\newcommand{\mbloc}[1]{\tiny{\bra{\begin{array}{ccccccc}#1\end{array}}}}

\begin{document}

\widetext


\title{Exact computation of the Maximum Entropy Potential of spiking neural networks models}
\author{R.~Cofr\'{e}} \affiliation{NeuroMathComp team (INRIA, UNSA LJAD) 2004 Route des Lucioles, 06902 Sophia-Antipolis, France
}
\author{B.~Cessac} \affiliation{NeuroMathComp team (INRIA, UNSA LJAD) 2004 Route des Lucioles, 06902 Sophia-Antipolis, France
}

\vskip 0.25cm
\date{\today}

\begin{abstract}
%
Understanding how stimuli and synaptic connectivity influence the statistics of spike patterns in neural networks is a central question in computational neuroscience.
Maximum Entropy approach has been successfully used to characterize the statistical response of simultaneously recorded spiking neurons responding to stimuli. But, in spite of good performance in terms of prediction, the fitting parameters do not explain the underlying mechanistic causes of the observed correlations. On the other hand, mathematical models of spiking neurons (neuro-mimetic models) provide a probabilistic mapping between stimulus, network architecture and spike patterns in terms of conditional probabilities. In this paper we build an exact analytical mapping between neuro-mimetic and Maximum Entropy models. 
\end{abstract}

\maketitle

\section{I. INTRODUCTION}
 \vspace{-10pt}
The spike response of a neural network to external stimuli is largely conditioned by the stimulus itself, synaptic interactions and  neuronal network history \cite{gerstner-kistler:02b}. Understanding these dependences is a current challenge in neuroscience \cite{rieke-warland-etal:97}. Since spikes occur irregularly both within and over repeated trials \cite{shadlen:94}, it is reasonable to characterize spike trains using statistical methods and probabilistic descriptions. 

Among existing approaches to characterize spike train statistics, the Maximum Entropy Principle (MaxEnt) \cite{jaynes:57} has been successfully applied to data from the cortex and the retina \cite{schneidman-berry-etal:06,shlens-field-etal:06,marre-boustani-etal:09,ganmor-segev-etal:11a}.
It consists of fixing a set of constraints, determined as the empirical average of observables measured from spiking activity. Maximizing the statistical entropy, given those constraints, provides a unique probability called Gibbs distribution. The choice of constraints determines a ``model". Prominent examples are the Ising model \cite{schneidman-berry-etal:06,shlens-field-etal:06} where constraints are firing rates and probabilities that $2$ neurons fire at the same time, the Ganmor-Schneidman-Segev model \cite{ganmor-segev-etal:11a}, which considers additionally the probability of triplets and quadruplets of spikes, or the Tka\v{c}ik et al model \cite{tkacik-marre-etal:13} where the probability that $K$ out of $N$ cells in the network generate simultaneous action potentials is an additional constraint. In these examples the statistics has no memory and successive times are considered statistically independent; but Markovian models where the probability of a spike pattern at a given time depends on the spike history can be considered as well \cite{marre-boustani-etal:09,vasquez-marre-etal:12}. MaxEnt models depends on a set of parameters (Lagrange multipliers) which are fitting parameters. In statistical physics language, these are parameters conjugated to the constraints, just like the inverse temperature $\beta=\frac{1}{k T}$ is conjugated to the energy, or chemical potential is conjugated with the number of particles. However, whereas inverse temperature or chemical potential have a clear interpretation thanks to the links between thermodynamics and statistical physics, the fitting parameters (Lagrange multipliers) used for spike train statistics do not benefit from such deep relations and are interpreted e.g. via loose analogies with magnetic systems. For example the Lagrange multipliers $\J_{ij}$ conjugated to pairwise spike coincidence are interpreted as ``functional interactions" \cite{ganmor-segev-etal:11a} due to their analogy with magnetic interactions in the Ising model. Likewise the parameters $\h_i$ conjugated with single spike events (whose average is the firing rate) are believed to be related with an effective stimulus received by neuron $i$. However, the connection between ``functional" interactions $\J_{ij}$ and real interactions (e.g synapses) in the network remains elusive, as well as the link between effective stimuli and the stimulus viewed by a neuron.

\begin{figure}[h]
\begin{center}
\scalebox{0.57}
{
\includegraphics{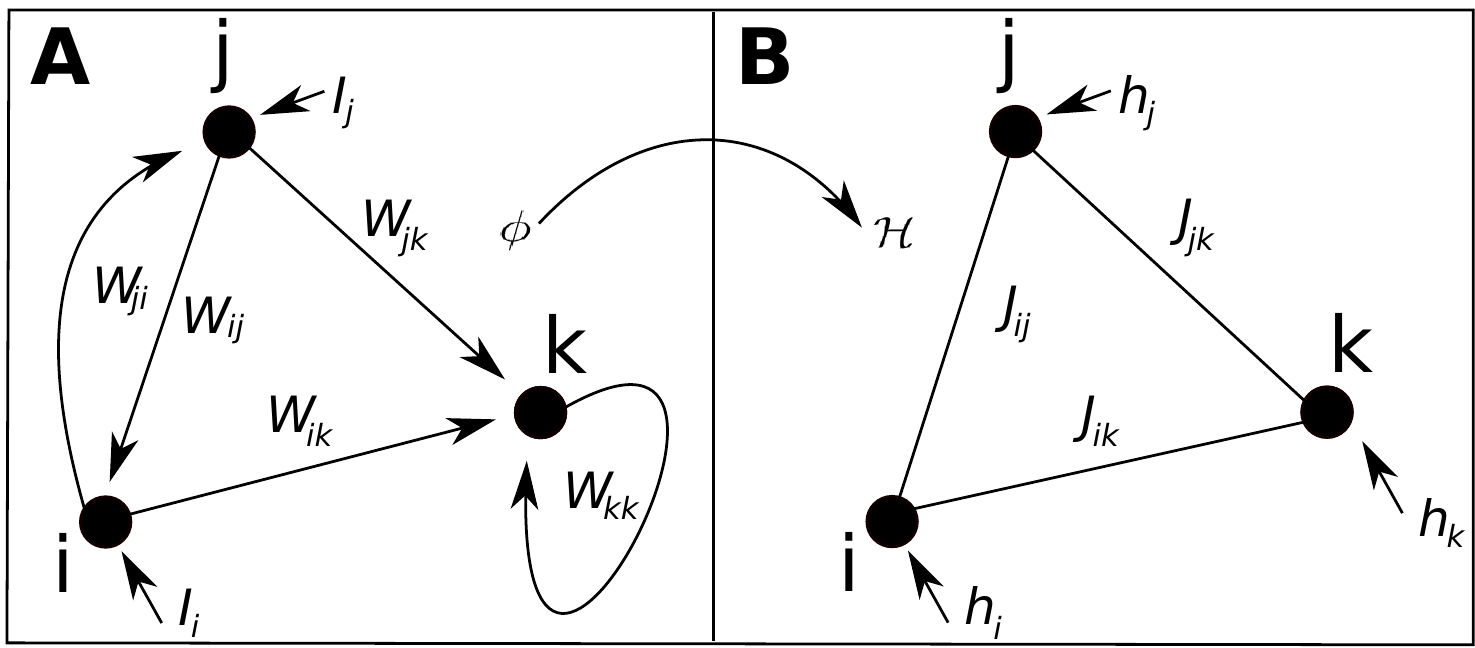}}
\caption{\label{fig:dual} \footnotesize{(A) Neuro-mimetic approach. Neurons are interacting via synaptic weights $W_{ij}$ and submitted to a stimulus $I$. Spike probabilities are explicit functions of these parameters. (B) MaxEnt statistical approach. Here the relation between neurons are expressed by functional parameters allowing to correctly fit the correlations in the model. The graph represents the Ising model where only local fields and pairwise interactions are drawn. More general interactions are considered in the text. In Ising model pairwise interactions are symmetric (represented without arrows). We are looking for an explicit and exact correspondence between these two representations.}}
\end{center}
 \vspace{-15pt}
\end{figure}

An alternative approach is based on spiking neuron models, providing a mathematical description of neural dynamics. These models give a probabilistic mapping between network architecture, stimuli, spiking history of the network and spiking response in terms of conditional probabilities of spike pattern given the network history.  Prominent examples are the Linear-Non Linear model (LN), the Generalized Linear Model (GLM) \cite{brillinger:88,chichilnisky:01} or  Integrate-and-Fire models \cite{gerstner-kistler:02b}. In all these examples the conditional probabilities that a spike pattern occurs at time $t$ given the network spike history are explicit functions of ``structural" parameters in the neural network (that can be interpreted as synaptic weights  $\mathcal{W}$ matrix, and stimulus $\mathcal{I}$) (Fig. \ref{fig:dual}a).

These conditional probabilities define a Markov process that mimics the biophysical dynamics of neurons in a network and the mechanisms that govern spike trains emission, including stimulus dependence and neurons interactions via synapses. We call them \textit{neuro-mimetic} statistical models.

To summarize, at least two different representations can be used to analyze spike train statistics in neural networks (fig. \ref{fig:dual}). The goal of this paper is to establish an explicit and exact correspondence between these two representations.\\
 
A previous result attempting to describe such a relationship can be found in \cite{cocco-leibler-etal:09}. Here, the authors fit a leaky Integrate and Fire model matching spike train data from a population of retinal ganglion cells. At the same time they fit a MaxEnt Ising model from this data. This allows them to compare in particular synaptic weights $W_{ij}$ with MaxEnt Ising couplings $\J_{ij}$. Another work in this direction can be found in \cite{granot-etal:13} in which stimulus dependent MaxEnt is introduced based on (LN) model, attempting to include stimulus information into the ``local fields" of the Ising model. Both examples are limited to the Ising model, thus do not include memory effects in the MaxEnt statistics.\\

We propose here a generalization which allows us to handle more general types of neuro-mimetic models as well as general spatio-temporal MaxEnt distributions (including memory). The method we used is based on Hammersley-Clifford decomposition \cite{hammersley-clifford:71} and periodic orbit invariance from ergodic theory \cite{pollicott-weiss:03}. The techniques are therefore different from \cite{cocco-leibler-etal:09,granot-etal:13}.\\

More generally, we answer the following questions:\\

\textit{ \textbf{Question 1:} Given an ergodic Markov process, where the transition probabilities are known, can we construct a MaxEnt potential, with a minimum of constraints, reproducing \underline{exactly} the (spatio-temporal) statistics of this process?}\\

This is the most general question we answer in this paper. It is important to notice that our results are not restricted to spike trains and neural networks, but to any ergodic Markov chain. The next question focuses on the correspondence between MaxEnt parameters and structural parameters defining spiking neural networks.\\

\textit{ \textbf{Question 2:} Given the transition probabilities from a neuro-mimetic model, is it possible to derive an analytic correspondence between MaxEnt fitting parameters (Lagrange multipliers)  and neuro-mimetic structural parameters? For example can we establish a correspondence between the local fields $h_i$ and the external stimulus $I_i$? Between Ising couplings $\J_{ij}$ with $W_{ij}$, the synaptic weights?}\\

We show that there exists an exact and analytic correspondence revealing that Lagrange multipliers are complex non-linear functions of structural parameters. For example the local field $\h_i$ or the functional interactions $\J_{ij}$ are non-linear functions depending \textit{generically} on \underline{all} stimuli and \underline{all} synaptic weights.\\

Additionally this correspondence raises up the question of dimensionality. A neuro-mimetic model with $N$ neurons typically has $N^2+N$ independent parameters ($N^2$ synaptic weights and $N$ stimuli); a MaxEnt model with memory depth $D$ may have up to $2^{N(D+1)}$ independent parameters (see text). The dimensionality of these two types of models is drastically different. When mapping a MaxEnt model to a neuro-mimetic model there is clearly a loss of dimensionality. \\



\textit{ \textbf{Question 3:} Consider a MaxEnt model equivalent to a neuro-mimetic model. Then the difference in dimensionality between the spaces of parameters of both models suggests that either many of MaxEnt parameters are zero, or that there are relations among them, i.e. they are not independent. What is the generic situation?}\\


The paper is organized as follows. In section II we introduce some notations and present MaxEnt models with spatio-temporal constraints. The introduction of memory requires to define the Gibbs distribution in a more general setting than usual statistical physics definition. Although this formalism is well known \cite{georgii:88}, it allows us to make the connection between Gibbs distributions and Markov chains, a necessary step in our construction. In section III we develop our method, based on equivalence between potentials, Hammersley-Clifford hierarchy and periodic orbits invariance.  In section IV we present an example based on a discrete time Integrate and Fire model in which we compute explicitly the ``local fields" $\h_i$ and ``Ising couplings" $\J_{ij}$ as non linear functions of $\mathcal{W},\mathcal{I}$. We finally present the conclusions of this work in which we address especially the issue of the difference of dimensionality between the space of parameters of MaxEnt and neuro-mimetic models.

\section{II. SETTING}
\vspace{-10pt}

We study a network composed by $N$ neurons. Time has been discretized so that a neuron can at most fire
a spike within one time bin. A spike train is represented by a binary
time series with entries $\omega_k(n)=1$ if  neuron $k$ fires at time $n$ and $\omega_k(n)=0$ otherwise. 
The \textit{spiking pattern} at time $n$ is the vector  $\omega(n) = \bra{\omega_k(n)}_{k=1}^{N}$.
A {\em spike block} $\bloc{n_1}{n_2}$ is an ordered list of spiking patterns
where spike times range from $n_1$ to $n_2$. 

A {\em potential} $\H$ of range $R=D+1$ is a function that associates
to each spike block $\bloc{0}{D}$ a real value. We assume $\H(\bloc{0}{D}) > - \infty$. Any such potential can be written:
\beq\label{Hh}
\H(\bloc{0}{D}) \, =\sum_{l=0}^{{L}} h_l \, m_l(\bloc{0}{D}),
\eeq
where $L=2^{NR}-1$. The $h_l$'s are real numbers whereas the function $m_l$ with $m_l(\bloc{0}{D})=\prod_{u=1}^r \omega_{k_u}(n_u)$ is called a \textit{monomial}. $m_l(\bloc{0}{D})=1$ if and only if neurons $k_u$ fires at time $n_u$ and zero otherwise. $r$ is the \textit{degree} of the monomial. By analogy with spin systems, we see from (\ref{Hh}) that monomials somewhat constitute formal \textit{spatio-temporal} interactions: degree $1$ monomials corresponds to ``self-interactions", degree $2$ to pairwise interactions, and so on.  The $h_l$'s characterize the intensity of the corresponding interaction. 

In many examples most $h_l$'s are equal to zero. For instance, Ising model considers
only monomials of the form $\omega_i(0)$ (singlets) or $\omega_i(0)\omega_j(0)$ (instantaneous pairwise events). 

\beq\label{Ising}
\H(\bloc{0}{D}) \, =\sum_{i=0}^N \h_i \omega_i(0) +  \, \sum_{i=0}^N \sum_{j >i}^N \J_{ij} \omega_i(0)\omega_j(0).
\eeq

More generally, the potential (\ref{Hh}) considers spatio-temporal events occurring within a time horizon $R$. 
This potential defines a unique \underline{stationary} probability $\mu$, called \textit{equilibrium state with potential $\H$} satisfying the following variational principle  \cite{georgii:88,Keller:98}:
\beq\label{VarPrinc}
\p{\H}=\sup_{\nu \in \cM} \pare{\s{\nu} \, + \, \noy{\H}}=
\s{\mu} \, + \, \moy{\H},
\eeq
\\
where $\cM$ is the set of stationary probabilities defined on the set of spike trains, whereas:

$$\s{\nu} \, = \, - \, \lim_{n \to \infty} \frac{1}{n+1} \, \sum_{\bloc{0}{n}} \, \noy{\bloc{0}{n}} \, \log \noy{\bloc{0}{n}},$$
\\
is the entropy of the probability $\nu$. The average of $\H$ with respect to $\nu$ is noted $\noy{\H}=\sum_{l=0}^L h_l \noy{m_l}$. In (\ref{VarPrinc}) we denote $\mu(\H), (\nu(\H))$ the average of $\H$ with respect to $\mu,(\nu)$. We insist on a point: we only consider stationary (time translation invariant) probabilities in this paper. 

The quantity $\p{\H}$ is called \textit{free energy} and has the following  properties:

\begin{itemize}

\item $\p{\H}$ is a log generating function of cumulants. We have:

$$\frac{\partial \p{\H}}{\partial h_l} =\mu\bra{m_l},$$
 the average of $m_l$ with respect to $\mu$ and:
\beq\label{d2Pbeta}
\frac{\partial^2 \p{\H}}{\partial h_k \partial h_l} = \frac{\partial \moy{m_l}}{\partial h_k} =
\sum_{n=-\infty}^{+\infty} C_{m_k\,,m_l}(n),
\eeq
where $C_{m_k\,,m_l}(n)$
%
%
is  the correlation function between the two monomials $m_k$ and $m_l$ at time $n$ in the equilibrium state $\mu$.
Note that correlation functions decay exponentially fast whenever $\H$ has finite range and $\H >-\infty$, thus $\sum_{n=-\infty}^{+\infty} C_{m_k\,,m_l}(n) < +\infty$.
Eq. (\ref{d2Pbeta}) characterizes the variation in the average value of $m_l$ when varying $h_k$ (linear response).
The corresponding matrix is a susceptibility matrix. It controls the  Gaussian fluctuations of monomials around their mean (central limit theorem)
 \cite{bowen:08}.  When considering potential of range 1 $(D=0)$ eq (\ref{d2Pbeta}) reduces to the classical fluctuation-dissipation theorem, because the corresponding process has no memory (successive times are independent thus $C_{m_k, m_l}(n)=0$ unless $n=0$).
 
 \item 
$\cP(\H)$ is a convex function of $h_l$'s. This ensures the uniqueness of the solution of (\ref{VarPrinc}).
\end{itemize}

\ssu{Transfer Matrix}

We now show that any potential $\H > -\infty$ with the form (\ref{Hh}) is naturally associated with a Markov chain whose invariant distribution is the equilibrium state $\mu$ satisfying the variational principle (\ref{VarPrinc}). $\mu$ is additionally a Gibbs distribution.\\

We first recall that a Markov chain is defined by a set of transition probabilities $\Probc{\omega(n)}{\bloc{n-D}{n-1}}$. We assume here that the memory depth of the chain $D$ is constant and finite, although an extension of the present formalism to variable length Markov chains ($D$ variable) or chains with complete connections ($D$ infinite) is possible \cite{fernandez-maillard:05}. We also assume that the chain is homogeneous (transition probabilities do not depend explicitly on time) and primitive (there exist $n$ such that any two states are connected by a path of length $n$, with positive probability) \footnote{In the present context this property is ensured by the assumption $\H > -\infty$ (sufficient condition).}. Then the Markov chain admits a unique invariant probability $\mu$ which obeys the Chapman-Kolmogorov relation: $\forall n_1,n_2, n_2 +D-1 > n_1, $

\beq\label{Chk}
\moy{\bloc{n_1}{n_2}} = \prod_{n=n_1}^{n_2-D} \Probc{\omega(n+D)}{\bloc{n}{n+D-1}} \moy{\bloc{n_1}{n_1+D-1}}.
\eeq

Introducing:

\beq\label{Phi2}
\phi (\bloc{n}{n+D}) = \log \Probc{\omega(n+D)}{\bloc{n}{n+D-1}},
\eeq

we have 

$$\moy{\bloc{n_1}{n_2}} = e^{\sum_{n=n_1}^{n_2-D} \phi(\bloc{n}{n+D})} \moy{\bloc{n_1}{n_1+D-1}}.$$
\\
$\phi$ is called a \textit{normalized potential}. To each $\H$ of the form (\ref{Hh}) corresponds a unique normalized potential $\phi$ and a unique invariant measure $\mu$. Although this correspondence can be found in many text-books (see for example \cite{georgii:88,Keller:98}), we summarize it here since it is the core of our approach.\\

\ssu{From $\H$ to Markov chains}

Each spike block is associated to a unique integer (index) $l =\sum_{k=1}^N \sum_{n=0}^D 2^{n\,N+k-1} \, \omega_k(n)$, where neurons $k=1,..,N$ are considered from top to bottom and time $n=0,..,D$ from left to right in the spike train. We denote $\om{l}$ the spike block corresponding to the index $l$. Here an example with $N=2$ and $R=3$, $\om{6}=\mbloc{0 & 1 & 0\\1 & 0 & 0}$. 

Consider two spike blocks $\om{l},\om{l'}$ of range $D\geq 1$. The transition $\om{l} \to \om{l'}$ is \textit{legal} if $\om{l},\om{l'}$ have a common block  $\bloc{1}{D-1}$.  Here is an example of a legal transition: 
$$
\tiny{\om{l} =\left[
\begin{array}{ccc}
0&0&1\\
0&1&1\\
\end{array}
\right] 
};
\,
\tiny{\om{l'} =\left[
\begin{array}{ccc}
0&1&1\\
1&1&0\\
\end{array}
\right]
}
.$$
and a forbidden transition:
$$
\tiny{\om{l} =\left[
\begin{array}{ccc}
0&0&1\\
0&1&1\\
\end{array}
\right] 
};
\,
\tiny{\om{l'} =\left[
\begin{array}{ccc}
0&1&1\\
0&1&0\\
\end{array}
\right]
}
.$$
Any block $\bloc{0}{D}$ of range $R=D+1$ can be viewed as a legal transition from the block $\om{l}=\bloc{0}{D-1}$ to the block 
$\om{l'}=\bloc{1}{D}$ and in this case we write $\bloc{0}{D} \sim \om{l}\om{l'}$.  \\

We construct the \textit{transfer matrix} $\cL$:
%
$$
\cL_{\om{l},\om{l'}}=   
\left\{
\begin{array}{lll}
 e^{\H(\bloc{0}{D})}
\quad &\mbox{if }  \bloc{0}{D} \sim \om{l}\om{l'}   \\
0, \quad &\mbox{otherwise}.
\end{array}
\right. 
$$
\\
This is a $2^{ND} \times 2^{ND}$ matrix whose indexes are spike blocks. Note that, from the assumption $\H > -\infty$, each legal transition corresponds to a positive entry in the matrix $\cL$. Therefore $\cL$ is primitive and satisfies the Perron-Frobenius theorem \cite{gantmacher:59}.

As a consequence of the Perron-Frobenius theorem, $\cL$ has a unique real positive eigenvalue $s$, strictly larger in modulus than the other eigenvalues, and with right, $\rpf$, and left, $\lpf$, eigenvectors: $\cL\rpf=s\rpf, \, \lpf\cL=s \lpf$.
%
The following holds:
\begin{enumerate}[(a)]
\item These eigenvectors have strictly positive  entries $\rpfc{\cdot}>0$, $\lpfc{\cdot}>0$; their arguments are blocks of range $D$. They can be chosen so that the scalar product 
$\langle L,R \rangle =1$. 
%
\item The following potential: 
\beq\label{Phi_norm}
\phi(\bloc{0}{D})=\H(\bloc{0}{D}) - \log \rpfc{\bloc{0}{D-1}} + \log \rpfc{\bloc{1}{D}}-\log s
\eeq
is normalized i.e. it defines via (\ref{Phi2}) an homogeneous Markov chain with transition probability $\Probc{\omega(D)}{\bloc{0}{D-1}} = e^{\phi(\bloc{0}{D})}$.

\item The unique invariant probability of this Markov chain is:
\beq\label{mu}
\mu(\bloc{0}{D-1})=\rpfc{\bloc{0}{D-1}}\lpfc{\bloc{0}{D-1}}.
\eeq

\item  It follows from Chapman-Kolmogorov equation (\ref{Chk}) and from (\ref{Phi_norm},\ref{mu}) that, for $D>0$:
\beq\label{fchc}
\moy{\bloc{0}{n}} = \frac{e^{\sum_{k=0}^{n-D}\H\pare{\bloc{k}{k+D}}}}{s^{n-D+1}} 
\rpfc{\bloc{n-D+1}{n}}\lpfc{\bloc{0}{D-1}}.
\eeq

\item $\mu$ obeys the variational principle (\ref{VarPrinc}) and 
$$\mathcal{P}[\mathcal{H}] =\log s.$$

When considering a normalized potential, the transfer matrix becomes a stochastic transition matrix with maximal eigenvalue 1. Thus $\mathcal{P}[\phi]=0.$ 
\item It follows from (\ref{fchc}) that 

$\exists \, A, B >0$ such that, for any block $\bloc{0}{n}$ the Gibbs distribution reads \cite{bowen:08,Keller:98}:
\beq\label{gibbs}
A \leq \frac{\moy{\bloc{0}{n}}}{e^{-(n-D+1)\cP(\H)} e^{-\sum_{k=0}^{n-D}\H\pare{\bloc{k}{k+D}}}}\leq B.
\eeq
%
%
This is actually the definition of Gibbs distributions in ergodic theory \footnote{When considering finite range potentials equilibrium states and Gibbs distributions are equivalent notions. This equivalence requires additional assumptions for infinite range potentials}.

\end{enumerate}

This definition encompasses the classical definition of Gibbs distributions, $\frac{e^{\mathcal{H}}}{Z}$ found in standard textbooks of statistical physics. Let us indeed consider a potential of range $R=1, (D=0)$. This is a limit case in the definition of the transfer matrix where transitions between spike patterns $\omega(0) \rightarrow \omega(1)$ are considered and where all transitions are legal. $\mathcal{L}_{\omega(0),\omega(1)} = e^{\mathcal{H}(\omega(0))}$, thus each row has the form:
$$(e^{\mathcal{H}(\omega(0))},e^{\mathcal{H}(\omega(0))},\hdots,e^{\mathcal{H}(\omega(0))}).$$
The matrix $\mathcal{L}$ is degenerated with a maximum eigenvalue:
$$s=Z=\sum_{\omega(0)} e^{\mathcal{H}(\omega(0))}$$
\\
and all other eigenvalues $0$. The left eigenvector corresponding to $s=Z$ is:
$$L=(\frac{1}{Z},\frac{1}{Z},\hdots,\frac{1}{Z})$$
\\
whereas $R(\omega(0))=e^{\mathcal{H}(\omega(0))}$. Note that we have normalized $L$ so that $\langle L,R \rangle =1$. We have therefore $\mu(\omega(0))=\frac{e^{\mathcal{H}(\omega(0))}}{Z}$, the classical form for the Gibbs distribution. The normalized potential in the limiting case is $\phi(\omega(0))=\mathcal{H}(\omega(0)) - \log Z$, whereas the Markov chain has no memory: successive events are independent. This last remark reflects a central weakness of memory-less MaxEnt models to describe neuron dynamics. They neither involve memory nor time causality. In order to consider realistic situations, where a spike pattern probability depends on the past spike history, MaxEnt models must be constructed as we did \footnote{Alternative constructions such as \cite{marre-boustani-etal:09} has been proposed, but they require additional assumptions such as detailed balance}. Indeed, it is not possible to extend the form $\frac{e^{\mathcal{H}(\omega(0))}}{Z}$ to spatio-temporal potentials. This is obvious from equation (\ref{fchc}): $\mu[\omega_0^n]$ has not the form $\frac{e^{\mathcal{H}(\omega(0))}}{Z_n}$ with $Z_n=\sum_{\omega_0^n} e^{\mathcal{H}(\omega_0^n)}$. This can also be readily seen from (\ref{gibbs}). However the following holds: 
$$\mathcal{P}[\mathcal{H}]= \lim_{n \rightarrow \infty} \frac{1}{n} \log Z_n.$$

This outlines a crucial point: as soon as one introduces memory in the MaxEnt, infinite time limits have to be considered in order to fully characterize the statistics. This is \textit{mutatis mutandis} the same procedure as taking the thermodynamic limit in spatial lattices \cite{georgii:88}.

\ssu{Equivalent potentials}\label{Eqp}

Although a potential $\mathcal{H}$ of the form (\ref{Hh}) corresponds (if $\mathcal{H}>-\infty$) to a unique normalized potential $\phi$ and Gibbs distribution $\mu$, this correspondence is not one to one. To a normalized potential $\phi$ corresponds infinitely many potentials of the form (\ref{Hh}). Hence two potentials $\mathcal{H}^{(1)},\mathcal{H}^{(2)}$ can correspond to the same Gibbs distribution (We call them \underline{equivalent}). \\
A standard result in ergodic theory states that  $\H^{(1)}$ and $\H^{(2)}$ are equivalent if and only if there exists a range $D>0$ function $f$ such that \cite{bowen:08}:
\begin{equation}\label{Cohomology}
\H^{(2)}\pare{\bloc{0}{D}}=\H^{(1)}\pare{\bloc{0}{D}} - f\pare{\bloc{0}{D-1}}+ f\pare{\bloc{1}{D}} + \Delta,
\end{equation} 
where $\Delta =\p{\H^{(2)}} - \p{\H^{(1)}}$. This relation establishes a \textit{strict} equivalence and does not correspond e.g. to renormalization. The validity of (\ref{Cohomology}) can be readily seen by plugging $\H^{(2)}$ in the variational formula (\ref{VarPrinc}) the terms corresponding to $f$ cancels because $\nu$ is time-translation invariant. Therefore, the supremum is reached for the same Gibbs distribution as $\H^{(1)}$ whereas $\Delta$ is indeed the difference of free energies. The ``only if" part is more tricky.

Equation (\ref{Phi_norm}) is a particular case of equation (\ref{Cohomology}), where $\H^{(2)}=\phi, \H^{(1)}=\H, f=\log R$ and $\Delta=-\log s$. This equation has the virtue to unify two very different approaches. It establishes a relation between Markov chain normalized potentials (\ref{Phi2}) on one hand and potentials of the form (\ref{Hh}) on the other hand (the arrow $\phi   \rightarrow \H$ in fig. \ref{fig:dual}). Equation (\ref{Cohomology}) answers therefore the first part of the question (1), but, by itself does not provide a straightforward way to exploit it, due to the arbitrariness in the choice of $f$. Indeed, there are infinitely many potentials $\H$ corresponding to the same Gibbs distribution (the same normalized potential $\phi)$.

This arbitrariness in the choice of $f$ raises a natural question closely related to the second part of question (1). Given a normalized potential is it possible to find, among the infinite family of equivalent potentials, a canonical form of $\H$ with a minimal number of terms ? The situation is a bit like normal forms in bifurcations theory where variable changes allows one to eliminate locally non resonant terms in the Taylor expansion of the vector field \cite{arnold:83}. Here, the role of the variable changes is played by $f$. By suitable choices of $f$ one should be able to eliminate some monomials in the expansion (\ref{Hh}). An evident situation corresponds to monomials related by time translation, e.g. $\omega_i(0)$ and $\omega_i(1)$: since any $\nu \in \cM$ is time translation invariant $\noy{\omega_i(0)}=\noy{\omega_i(1)}$, the firing rate of neuron $i$. Such monomials correspond to the same constraint in (\ref{VarPrinc}) and can therefore be eliminated.  A potential where monomials, related by time translation, have been eliminated (the corresponding $h_l$ vanishes) is called \textit{canonical}. A canonical potential contains thus, in general, $2^{NR} - 2^{N(R-1)}$ terms. We now show that canonical potentials cannot be further reduced.

\ssu{Canonical interactions cannot be eliminated using the equivalence equation (\ref{Cohomology})}

Assume  that we are given two potentials $\H^{(1)}, \H^{(2)}$ in the \underline{canonical form},
where $\H^{(1)}$ has a zero coefficient for the canonical interaction $m_l$ whereas
$\H^{(2)} = \H^{(1)} + h_l \, m_l$, $h_l \neq 0$. Let us show that these two potentials
are not equivalent. For this we need to introduce a bit of notations further used in the text.
\\
Since a monomial is defined by a set of spike events $(k_u, n_u)$, one can associate to each monomial a spike block or ``mask" where the only bits '$1$' are located at $(k_u,n_u)$, $u=1, \dots, r$. This mask has therefore an index.
Whereas the labeling of monomials in (\ref{Hh}) was arbitrary, $m_l$ denotes from
now on the monomial with mask $\om{l}$. Let us define the block \textit{inclusion} $\sqsubseteq$, by
$\bloc{0}{D} \sqsubseteq \seq{\omega'}{0}{D}$ if $\omega_k(n)=1 \Rightarrow \omega'_k(n)=1$, with the convention
that the block of degree $0$, $\om{0}$, is included in all blocks. 
Then, for two integers $l,l'$:

\beq \label{inc}
m_{l'}(\om{l})=1 \mbox{ if and only if } \om{l'}  \sqsubseteq \om{l}.
\eeq 
\\
Now, from (\ref{Cohomology}), $\H^{(2)} = \H^{(1)} + h_l \, m_l$ and $\H^{(1)}$  are therefore equivalent if one can find a $D$-dimensional function $f$ such that, $\forall \bloc{0}{D}$:

$$
f\pare{\bloc{1}{D}}- f\pare{\bloc{0}{D-1}} + \Delta + h_l \un_{\omega^{(l)} \sqsubseteq \bloc{0}{D}}=0,
$$
\\
where $\un_{\omega^{(l)} \sqsubseteq \bloc{0}{D}}$ is the standard indicator function that takes value 1 when $\omega^{(l)} \sqsubseteq \bloc{0}{D}$ and 0 otherwise. Let us consider 2 specific blocks. The block only composed by '1's contains all other blocks, and it is translation invariant so that the terms involving $f$ cancel in the equation above. We have therefore $ \Delta + h_l=0$. The block only composed by '0's is also translation invariant and, if $l>0$
we obtain $\Delta=0$, so that $h_l=0$, in contradiction with the hypothesis. Therefore, two canonical potentials are equivalent \textit{if and only if} all their canonical their coefficients $h_l$, $l>0$, are equal \cite{cessac-cofre:13c}.\\

There is still an arbitrariness due to the term $h_0$ (``Gauge" invariance). One can set it equal $0$ without loss of generality. In this way, there is only one canonical potential, with a minimal number of monomials, corresponding to a given stationary Markov chain.  

In the next section we show how to compute the coefficients of the canonical potential $\H^{(2)}$ equivalent to a known $\H^{(1)}$.

 \section{III. METHOD}

 \vspace{-10pt}
 
Given a spike block $\om{l_0}$, a \textit{periodic orbit} of period $\kappa$ is a sequence of spike blocks $\om{l_n}$ where $\om{l_{p \kappa +  n}} = \om{l_n}$, $p \geq 0$, $0 \leq n \leq \kappa -1$. From equation (\ref{Cohomology}) we have, for such a periodic orbit, 

\beq \label{Li}
\sum_{n=0}^{\kappa-1} \H^{(2)}\pare{\om{l_n}}= \sum_{n=0}^{\kappa-1} \H^{(1)}\pare{\om{l_n}} + \kappa \Delta,
\eeq 
\\
because the $f$-terms disappear when summed along a periodic orbit. It follows that the sum of a potential along a periodic orbit is an invariant (up to the constant term $\kappa \Delta$) in the class of equivalent potentials. This is a classical result in ergodic theory extending to infinite range potentials \cite{livsic:72}. This equation is valid whatever periodic orbit is considered. 
It is singularly useful if one takes advantage of an existing hierarchy between blocks and between monomials, the Hammersley-Clifford (H-C) hierarchy \cite{hammersley-clifford:71} that we explain now.

\ssu{Hammersley-Clifford hierarchy}

The construction of our method is based on the (H-C) factorization theorem, proved in the seminal although unpublished paper \cite{hammersley-clifford:71}. Later simpler proofs were given in \cite{Grimmett:73, Besag:74}. This result establishes the equivalence between Markov random fields and Gibbs distributions. It was proved in the context of undirected graphs where the clique structures provide the factorization of the potential. Our result is based on a decomposition of the potential over inclusions $\sqsubseteq$ of (spatio-temporal) blocks defined previously. Inclusions provide a hierarchical structure similar to the blackening algebra of (H-C). However (H-C) theorem does not provide by itself an explicit method to obtain from a Markov chain the corresponding  canonical MaxEnt potentials. On the opposite, our method of periodic orbits allows to perform this computation.

We can express $\H^{(2)}$ in the form (\ref{Hh}), then using (\ref{inc}) it follows that (\ref{Li}) becomes: 
\beq \label{Livsic2}
\sum_{n=0}^{\kappa-1} \sum_{l \sqsubseteq l_n} h_l^{(2)}  \, m_l(\om{l_n})= \sum_{n=0}^{\kappa-1} \H^{(1)} (\om{l_n})  + \kappa \Delta
\eeq
where, with a slight abuse of notations $l \sqsubseteq l_n$ stands for $\om{l}  \sqsubseteq \om{l_n}$.

The interesting fact about this representation is that the l.h.s of this equation is written entirely in terms of blocks included in the blocks considered in the periodic orbits. Therefore in order to compute all the coefficients $h_l$'s that characterize the canonical MaxEnt potential we can proceed by first obtaining the coefficient of degree 0, then the coefficients of degree 1,2, and so on. We use equation (\ref{Livsic2}) to compute from a known $\H^{(1)}$ potential its associated canonical potential $\H^{(2)}$. From now on we focus in the particular case when $\H^{1}=\phi$ is a normalized potential. To alleviate notation we note $\H^{(2)}=\H$.\\
\\
\textbf{Degree 0: Free energy}
\\
Start from the first mask in hierarchy, the mask $\om{0}$ containing only $0$'s, whose corresponding monomial is $m_0=1$ and consider its periodic orbit, of period $\kappa=1$, $\Set{\om{0}}$. The application of (\ref{Livsic2}) gives $h_0 = \phi (\om{0}) + \p{\H}$ and since we choose $h_0=0$ for the canonical potential we obtain a direct way to compute the free energy of $\H$.\\

\beq \label{tpres}
\p{\H}= - \phi (\om{0})
\eeq
\\
\textbf{Degree 1: Local Fields}
\\
Let us now consider masks of degree 1:
 $$\om{l_0}=\mbloc{0 & 0 & \cdots & 0\\ \vdots & \vdots & \cdots & \vdots \\ 0 & 0 & \cdots & 1 \\ \vdots & \vdots & \cdots & \vdots \\ 0 & 0 & \cdots & 0},$$
(where dots correspond to 0) corresponding to the monomial $\omega_i(D)$. Also consider the periodic orbit obtained by a $R$-circular shift of this block ($\kappa=R$): 

$$\om{l_0}=\mbloc{0 & 0 & \cdots & 0\\ \vdots & \vdots & \cdots & \vdots \\ 0 & 0 & \cdots & 1 \\ \vdots & \vdots & \cdots & \vdots \\ 0 & 0 & \cdots & 0} \to \om{l_1}=\mbloc{0 & \cdots & 0 & 0\\ \vdots & \cdots & \vdots & \vdots \\ 0 & \cdots & 1 & 0 \\ \vdots & \cdots & \vdots & \vdots \\ 0 & \cdots & 0 & 0}$$
\vspace{-5mm}
$$\quad \quad \quad \quad \quad \quad \quad \quad \cdots  \to \om{l_D}=\mbloc{0 & 0 & \cdots & 0\\ \vdots & \vdots & \cdots & \vdots \\ 1 & 0 & \cdots & 0 \\ \vdots & \vdots & \cdots & \vdots \\ 0 & 0 & \cdots & 0}.$$
\\
Since the corresponding monomials of the blocks in the orbit $\om{l_0},\om{l_1},\ldots,\om{l_D}$ are related by time translation they correspond to the same constraint in (\ref{VarPrinc}). The coefficient of all but one of these monomials is therefore set to $0$ in the canonical potential $\H$. We use the convention to keep the monomial $m_{l_0}$ whose mask contains a $1$ in the right most column. This convention extends to the monomials considered below. The block $\om{l}$ considered to generate this periodic orbit has one spike corresponding to neuron $i$. To make this explicit we note $h_l \equiv \h_i$: Then, applying (\ref{Livsic2}) to this periodic orbit we obtain: 

\begin{align}\label{h_l}
\h_i &= \phi(\om{l_0}) + \phi (\om{l_1}) + \cdots + \phi (\om{l_D}) \\
 & +R \phi (\om{0}) \nonumber
\end{align}

We have thus obtained the coefficient corresponding to the monomial $\omega_i(0)$ which is precisely $\h_i$ in Ising model (\ref{Ising}).

Considering a different $\om{l_0}$ of degree 1 and the periodic orbit generated by its $R$-circular shift we get another local field term. We do the same for the $N$ neurons.\\
\\
\textbf{Degree 2: Instantaneous pairwise interactions}
\\
Let us now consider instantaneous pairwise interactions.  We consider masks of the form :

$$\om{l_0}=\mbloc{0 & 0 & \cdots & 0\\ \vdots & \vdots & \cdots & \vdots \\ 0 & 0 & \cdots & 1 \\ \vdots & \vdots & \cdots & \vdots \\ 0 & 0 & \cdots & 1\\ \vdots & \vdots & \cdots & \vdots \\ 0 & 0 & \cdots & 0},$$
\\
corresponding to the monomial $\omega_i(D)\omega_j(D)$, the procedure is the same as above i.e. take the periodic orbit:
$$\om{l_0}=\mbloc{0 & 0 & \cdots & 0\\ \vdots & \vdots & \cdots & \vdots \\ 0 & 0 & \cdots & 1 \\ \vdots & \vdots & \cdots & \vdots \\ 0 & 0 & \cdots & 1\\ \vdots & \vdots & \cdots & \vdots \\ 0 & 0 & \cdots & 0}  \to \om{l_1}=\mbloc{0 & \cdots & 0 & 0\\ \vdots & \vdots & \vdots & \vdots \\ 0 & \cdots & 1 & 0 \\ \vdots & \vdots & \vdots & \vdots \\ 0 & \cdots & 1 & 0\\ \vdots & \vdots & \vdots & \vdots \\ 0 & \cdots & 0 & 0}$$
\vspace{-5mm}
$$\quad \quad \quad \quad \quad \quad \quad \quad  \cdots \to \om{l_D}=\mbloc{0 & 0 & \cdots & 0\\ \vdots & \vdots & \cdots & \vdots \\ 1 & 0 & \cdots & 0 \\ \vdots & \vdots & \cdots & \vdots \\ 1 & 0 & \cdots & 0\\ \vdots & \vdots & \cdots & \vdots \\ 0 & 0 & \cdots & 0}.$$
\\
The coefficients corresponding to this monomials are $\J_{ij}$ in the Ising model (\ref{Ising}). We have, from (\ref{Livsic2}):

\beq\label{Jijising}
\J_{ij}=
\sum_{n=0}^{R-1} \phi \pare{\om{\sigma^n l}} + R \phi(\omega^{(0)})
-
\sum_{n=0}^{R-1} \sum_{l'_n \sqsubset \sigma^n l} h_{l'_n}.
\eeq
For blocks $l'_n \sqsubset \sigma^n l$ of degree $1$ the spike is either on neuron $i$ or neuron $j$. The contribution of these blocks is $\h_i \, + \h_j$.
In the blocks $l'_n \sqsubset \sigma^n l$ there is also the block $\om{0}$, whose contribution is $h_0=0$. Therefore, we finally have:

\beq\label{Jijising2}
\J_{ij}=
\sum_{n=0}^{R-1} \phi \pare{\om{\sigma^n l}} 
 + R\phi(\omega^{(0)}) - \h_i - \h_j .
\eeq
\\
\textbf{Degree 2: (1 time-step memory)}:
\\
For the one step of memory pairwise coefficients (e.g. $\omega_j(0) \omega_i(1)$)
the situation is slightly different;
$$\om{l_0}=\mbloc{0 & \cdots & 0 & 0\\ \vdots & \vdots & \vdots & \vdots \\ 1 & 0 & \cdots & 0 \\ \vdots & \vdots & \vdots & \vdots \\ 0 & 1 & \cdots & 0\\ \vdots & \vdots & \vdots & \vdots \\ 0 & \cdots & 0 & 0},$$
Here the periodic orbit generated by the $R$-circular shift of $\om{l_0}$ is:

$$\om{l_0}=\mbloc{0 & \cdots & 0 & 0\\ \vdots & \vdots & \vdots & \vdots \\ 1 & 0 & \cdots & 0 \\ \vdots & \vdots & \vdots & \vdots \\ 0 & 1 & \cdots & 0\\ \vdots & \vdots & \vdots & \vdots \\ 0 & \cdots & 0 & 0} \to \om{l_1}=\mbloc{0 & \cdots & 0 & 0\\ \vdots & \vdots & \vdots & \vdots \\ 0 & \cdots & 0 & 1 \\ \vdots & \vdots & \vdots & \vdots \\ 1 & \cdots & 0 & 0\\ \vdots & \vdots & \vdots & \vdots \\ 0 & \cdots & 0 & 0}$$
\vspace{-5mm}
$$\om{l_2}=\mbloc{0 & \cdots & 0 & 0\\ \vdots & \vdots & \vdots & \vdots \\ 0 & \cdots & 1 & 0 \\ \vdots & \vdots & \vdots & \vdots \\ 0 & \cdots & 0 & 1\\ \vdots & \vdots & \vdots & \vdots \\ 0 & \cdots & 0 & 0} \to \cdots \to \om{l_D}=\mbloc{0 & 0 & 0 & \cdots\\ \vdots & \vdots & \vdots & \vdots \\ 0 & 1 & 0 & \cdots \\ \vdots & \vdots & \vdots & \vdots \\ 0 & 0 & 1 & \cdots\\ \vdots & \vdots & \vdots & \vdots \\ 0 & \cdots & 0 & \cdots}$$
\\
This orbit is not sufficient because it contains $2$ unknowns in eq (\ref{Livsic2}), namely the first and second blocks $\om{l_0}$ and $\om{l_1}$ correspond to monomials $\omega_j(D-1)\omega_i(D)$ and $\omega_j(D)\omega_i(0)$ which are not related by time translation, so correspond to different canonical constraints both having degree 2. Therefore it is not possible to solve (\ref{Livsic2}) just generating one circular periodic orbit. Fortunately is possible to circumvent this problem by generating additional periodic orbits. \\

Let us now consider the following periodic orbit:

$$\om{l_0}=\mbloc{0 & \cdots & 0 & 0\\ \vdots & \vdots & \vdots & \vdots \\ 1 & 0 & \cdots & 0 \\ \vdots & \vdots & \vdots & \vdots \\ 0 & 1 & \cdots & 0\\ \vdots & \vdots & \vdots & \vdots \\ 0 & \cdots & 0 & 0} \to \om{l_1}=\mbloc{0 & \cdots & 0 & 0\\ \vdots & \vdots & \vdots & \vdots \\ 0 & \cdots & 0 & 0 \\ \vdots & \vdots & \vdots & \vdots \\ 1 & \cdots & 0 & 0\\ \vdots & \vdots & \vdots & \vdots \\ 0 & \cdots & 0 & 0}$$
\vspace{-5mm}
$$\om{l_2}=\mbloc{0 & \cdots & 0 & 0\\ \vdots & \vdots & \vdots & \vdots \\ 0 & \cdots & 0 & 0 \\ \vdots & \vdots & \vdots & \vdots \\ 0 & \cdots & 0 & 0\\ \vdots & \vdots & \vdots & \vdots \\ 0 & \cdots & 0 & 0} \to \om{l_3}=\mbloc{0 & \cdots & 0 & 0\\ \vdots & \vdots & \vdots & \vdots \\ 0 & \cdots & 0 & 1 \\ \vdots & \vdots & \vdots & \vdots \\ 0 & \cdots & 0 & 0\\ \vdots & \vdots & \vdots & \vdots \\ 0 & \cdots & 0 & 0}$$
\vspace{-5mm}
\beq\label{p.o}
\om{l_4}=\mbloc{0 & \cdots & 0 & 0\\ \vdots & \vdots & \vdots & \vdots \\ 0 & \cdots & 1 & 0 \\ \vdots & \vdots & \vdots & \vdots \\ 0 & \cdots & 0 & 1\\ \vdots & \vdots & \vdots & \vdots \\ 0 & \cdots & 0 & 0} \cdots \to \om{l_{R+2}}=\mbloc{0 & 0 & 0 & \cdots\\ \vdots & \vdots & \vdots & \cdots \\ 0 & 1 & 0 & \vdots \\ \vdots & \vdots & \vdots & \cdots \\ 0 & 0 & 1 & \vdots\\ \vdots & \vdots & \vdots & \cdots \\ 0 & 0 & 0 & \cdots}
\eeq 
We have generated a periodic orbit, where (\ref{Livsic2}) has only one unknown, the coefficient associated to the first block. All the other blocks in the orbit are either of lower degree, thus we have already computed them; or are time translations of the first block, thus their coefficient is set to zero.

This is a particular example of a general procedure that we describe now. It allows to compute hierarchically \underline{any} $h_l$. The procedure is general, but we illustrate it with: 

$$\om{l}=\tiny{\bra{
\begin{array}{ccccccc}
0 & 0 & 1 & 1 & 0 & 1\\
0 & 0 & 1 & 0 & 1 & 1\\
\end{array}
}} $$

\begin{itemize}

\item Step 1. Shift circularly $\om{l}$ until the left-most spiking pattern has at least a $1$. Each of the circular shifts generate a mask, which corresponds to the same constraint in (\ref{VarPrinc}) so the corresponding $h_l$ coefficient is set to zero.

$$
\tiny{\bra{
\begin{array}{ccccccc}
0 & 0 & 1 & 1 & 0 & 1\\
0 & 0 & 1 & 0 & 1 & 1\\
\end{array}
}} 
\to
\tiny{\bra{
\begin{array}{ccccccc}
0 & 1 & 1 & 0 & 1 & 0\\
0 & 1 & 0 & 1 & 1 & 0\\
\end{array}
}} 
\to
\tiny{\bra{
\begin{array}{ccccccc}
1 & 1 & 0 & 1 & 0 & 0\\
1 & 0 & 1 & 1 & 0 & 0\\
\end{array}
}} 
$$

\item Step 2. Continue circularly left shifting but, before shifting, remove the $1$ with the lower neuron index, on the left most spike pattern. Tag the $1$s that has been removed. Do this until the total number of left shifts including step $1$ and $2$ is $R$.

$$
\quad \tiny{\bra{
\begin{array}{ccccccc}
1 & 0 & 1 & 0 & 0 & 0\\
0 & 1 & 1 & 0 & 0 & 1\\
\end{array}
}} 
\to
\tiny{\bra{
\begin{array}{ccccccc}
0 & 1 & 0 & 0 & 0 & 0\\
1 & 1 & 0 & 0 & 1 & 0\\
\end{array}
}} 
$$
\vspace{-6mm}
$$
\to
\tiny{\bra{
\begin{array}{ccccccc}
1 & 0 & 0 & 0 & 0 & 0\\
1 & 0 & 0 & 1 & 0 & 0\\
\end{array}
}} 
\to
\tiny{\bra{
\begin{array}{ccccccc}
0 & 0 & 0 & 0 & 0 & 0\\
0 & 0 & 1 & 0 & 0 & 1\\
\end{array}
}} 
$$

\item Step 3. Same as step 1. All the masks generated at this step correspond to the same constraint and thus have a zero coefficient.

$$
\tiny{\bra{
\begin{array}{ccccccc}
0 & 0 & 0 & 0 & 0 & 0\\
0 & 1 & 0 & 0 & 1 & 0\\
\end{array}
}} 
\to
\tiny{\bra{
\begin{array}{ccccccc}
0 & 0 & 0 & 0 & 0 & 0\\
1 & 0 & 0 & 1 & 0 & 0\\
\end{array}
}} 
$$

\item Step 4. Do the opposite of what was done in step 2: Restore the $1$'s that has been removed on the left most spike pattern while circularly shifting. In this way we finally regenerate $\om{l}$. 

$$
\quad \tiny{\bra{
\begin{array}{ccccccc}
0 & 0 & 0 & 0 & 0 & 1\\
0 & 0 & 1 & 0 & 0 & 1\\
\end{array}
}} 
\to
\tiny{\bra{
\begin{array}{ccccccc}
0 & 0 & 0 & 0 & 1 & 1\\
0 & 1 & 0 & 0 & 1 & 0\\
\end{array}
}} $$
\vspace{-6mm}
$$
\to
\tiny{\bra{
\begin{array}{ccccccc}
0 & 0 & 0 & 1 & 1 & 0\\
1 & 0 & 0 & 1 & 0 & 1\\
\end{array}
}} 
\to
\tiny{\bra{
\begin{array}{ccccccc}
0 & 0 & 1 & 1 & 0 & 1\\
0 & 0 & 1 & 0 & 1 & 1\\
\end{array}
}} 
$$

\end{itemize}
\quad 
\\
As claimed we have generated a periodic orbit where all monomials, but $\omega^{(l)}$, have either a coefficient 0 or have a degree smaller than $\omega^{(l)}$ and have therefore been already computed. Obviously, when getting to larger and larger degrees the method becomes rapidly intractable because of the exponential increase in the number of terms. The hope is that the influence of monomials decays rapidly with their  degree. Additionally, applying it to real data where transition probabilities are not exactly known leads to severe difficulties.  These aspects will be treated in a separated paper.
Our goal here was to answer the first question asked in the introduction. This goal is now achieved.\\ 
We now switch to the second question.

\ssu{From neuro-mimetic models to normalized potentials}\label{nmt}

In neuro-mimetic models the probability that the spike pattern $\omega(n)$ occurs at time $n$  
is the result of the complex membrane potentials dynamics \cite{gerstner-kistler:02b}. A simplification consists
of assuming that this probability is only a function of the spike history up to a certain memory
depth $D$. In this framework it is possible to consider a Markov chain in which the set of states consists of spike blocks $\omega_0^{D-1}$ from which legal transitions between blocks provides a family of conditional probabilities $\Probct{n}{\omega(n)}{\bloc{n-D}{n-1}}$ that may depend explicitly on time as indicated by the sub-index $n$. In neuro-mimetic models these probabilities depends on parameters that mimics biophysical quantities such as  synaptic weights matrix $\cW$ and stimulus $\mathcal{I}$.
A particularly important example is the Generalized Linear Model (GLM), which assimilates the spike response as
 an inhomogeneous point process, with
``conditional intensity" $\lambda_k(t|H_t)$ which modulates the probability that neuron $k$ emits a spike between times $t$ and $t+ dt$ given $H_t$ the history of spikes up to time $t$.  This function is given by \cite{pillow-ahmadian-etal:11}:
\beq\label{GLM2}
\lambda_k(t|H_t)=f\pare{b + K \cdot i(t) + \sum_{j} M_{kj} \cdot \omega_j},
\eeq
where $f$ is a non linear function; $b$ is a vector fixing the baseline firing rate of neurons; $K$ is a causal, time-translation invariant, linear convolution kernel that mimics a linear receptive field of neurons;  $i(t)$ is a stimulus;  $M_{kj}$ is a memory kernel that describes excitatory or inhibitory post spike effects of pre-synaptic neurons $j$ on post-synaptic neuron $k$. $\omega_j$ is the spike train of neuron $j$. Considering a discretization of time (time-steps $\Delta t$) and a memory cut-off (memory $D \rightarrow H^{n-1}_{n-D}$) from the conditional intensity we can get the conditional probability that neuron $k$ fire at time $n$:

$$P[\omega_k (n) = 1 | H^{n-1}_{n-D}] \approx \lambda_k (n | H_{n-D}^{n-1} )\Delta t := p_k (n).$$
\\
A crucial assumption in (GLM) is the conditional independence between neurons given the history $H_t$. Note that our formalism does not make this assumption and could be extended to neuro-mimetic models violating this assumptions (e.g models with gap junctions \cite{cofre-cessac:13}). Here for simplicity we stick at models with conditional independence. 
As a consequence of the conditional independence the probability of a spike pattern reads:

$$P[\omega(n)|\omega_{n-D}^{n-1}] \approx \prod_{k=1}^N p_k (n)^{\omega_k (n)} (1 - p_k (n))^{1-\omega_k (n)},$$
\\
and taking logarithm we get the normalized potential (\ref{Phi2}):

\begin{eqnarray}\label{PhiGLM}
    \phi(\bloc{n-D}{n}) &=& \sum_{k=1}^N \bigg[
\omega_k(n)\, \log p_k(n) \\ 
      & & \,+ \, \pare{1-\omega_k(n)}\log (1-p_k(n)) \bigg] \nonumber,
\end{eqnarray}
\\
We can use equation (\ref{Livsic2}) with $\phi=\H^{(1)}$ to compute from $\phi$ its canonical potential $\H^{(2)}=\H$:
\beq \label{Livsic}
\sum_{n=0}^{\kappa-1} \sum_{l \sqsubseteq l_n} h_l \, m_l(\om{l_n})= \sum_{n=0}^{\kappa-1} \phi \pare{\om{l_n}}  + \kappa \p{\H}.
\eeq 

Using this equation with the appropriate periodic orbits considered in the hierarchical order we obtain the corresponding canonical potential $\H$ from the normalized potential characterizing the GLM (\ref{PhiGLM}). We show in the following section using a different example (discrete time Leaky Integrate and Fire model) how the coefficients corresponding to firing rates, instantaneous and 1-step memory pairwise correlations can be calculated explicitly in terms of synaptic weights and stimulus.

\su{IV. The discrete time Leaky Integrate and Fire model}\label{SBMS}

In this section we illustrate our result in a stochastic leaky Integrate-and-Fire model with noise and stimulus \cite{soula-beslon-etal:06} analyzed rigorously in \cite{cessac:11a}. 

This model is a discretization of the usual leaky Integrate-and-Fire model. Its dynamics reads:
\begin{equation}\label{Bms}
V (t + 1) = F (V (t)) + \sigma_B B(t),
\end{equation}
where $V(t) = \pare{V_i(t)}_{i=1}^N$ is the vector of neuron's membrane potential at time $t$; $F(V)$ is a vector-valued function
with entries:
$$F_i (V) = \gamma V_i (1 - S\bra{V_i }) +
\sum_{j=1}^N W_{ij} S\bra{V_j } + I_i, \quad i=1 \dots N$$
where $\gamma \in [0, 1[,$ is the (discrete-time) ``leak rate \footnote{Thus, it corresponds to $\gamma = 1 -\frac{dt}{RC}$ in the continuous-time LIF model.}"; $S$ is a function characterizing the neuron's firing: for a firing threshold $\theta > 0$, $S(x) = 1$ whenever $x \geq \theta$ and $S(x) = 0$ otherwise; $I_i$ is an external current. In the most general version of this model, $I_i$ depends on time. Here, we focus on the case where $I_i$ is constant, ensuring the stationarity of dynamics.
Finally, in  (\ref{Bms}), $\sigma_B > 0$ is a variable controlling the noise intensity, where the vector $B(t) = (B_i(t))_{i=1}^N$ is an additive noise. It has Gaussian independent and identically distributed entries with zero mean and variance 1. 

\ssu{The normalized potential}

The normalized potential of the model (\ref{Bms}) has infinite range. Indeed, a neuron has memory only back to the last time when it has fired. But this time is unbounded (although the probability that the last firing time arises before time $m$ decays exponentially fast as $m \rightarrow -\infty$). Nevertheless, the exact potential can be approximated by the finite range potential \cite{cessac:11a}.

\begin{eqnarray}\label{PhiBMS}
    \phi(\bloc{0}{D}) &=& \sum_{k=1}^N \bigg[
\omega_k(D)\, \log\pi\pare{\X{k}{\bloc{0}{D-1}}} \\ 
      & & \,+ \, \pare{1-\omega_k(D)}\log\pare{1-\pi(\X{k}{\bloc{0}{D-1}}} \bigg] \nonumber,
\end{eqnarray}
\\
where the function $\pi$ is:
$$\pi(x) = \frac{1}{\sqrt{2\pi}}\int_{x}^{+\infty} e^{\frac{-u^2}{2}}du.
$$
$\phi$ has therefore the form of a (GLM) (\ref{PhiGLM}). All functions appearing below depend on the spike block $\bloc{0}{D-1}$ and make explicit the dependence of the network state (membrane potentials) on the spike history of the network. 

The term:
\beq\label{Xk}
\X{k}{\bloc{0}{D-1}}=\frac{\theta-\Vd{k}{\bloc{0}{D-1}}}{\sigma_k(\bloc{0}{D-1})},
\eeq
contains the network spike history dependence of the neuron $k$ at time $D$. More precisely, the term $\Vd{k}{\bloc{0}{D-1}}$ contains the deterministic part of the membrane potential of neuron $k$ at time $D$, given the network spike history $\bloc{0}{D-1}$, whereas $\sigma_k(\bloc{0}{D-1})$ characterizes the variance of the integrated noise in the  neuron $k$'s membrane potential. We have:
$$
\Vd{k}{\bloc{0}{D-1}}=\sum_{j=1}^N W_{kj} \, \et{kj}{\bloc{0}{D-1}}+I_k\frac{1-\gamma^{D-\tO{k}{\bloc{0}{D-1}}}}{1-\gamma}.
$$
The first term is the network contribution to the neuron $k$'s membrane potential, where:
$$
\et{kj}{\bloc{0}{D-1}}=\sum_{l=\tO{k}{\bloc{0}{D-1}}}^{D-1} \gamma^{D-1-l} \omega_j(l),
$$
is the sum of spikes emitted by $j$ in the past, with a weight $\gamma^{D-1-l}$ corresponding to the leak decay of the spike influence as time goes on. The notation $\tO{k}{\bloc{0}{D-1}}$ means the last time before $D-1$ where neuron $k$ has fired, with the convention that this time is $0$ if neuron $k$ didn't fire between $0$ and $D-1$ in the block $\bloc{0}{D-1}$. In the definition of $\et{kj}{\bloc{0}{D-1}}$ we sum from $\tO{k}{\bloc{0}{D-1}}$: this is because the membrane potential of neuron $k$ is reset whenever $k$ fires, hence loosing the memory of its past.
Finally, in (\ref{Xk}), we have:
$$ \sigma_k^2(\bloc{0}{D-1})=\sigma_B^2\frac{1-\gamma^{2 \left(D-\tO{k}{\bloc{0}{D-1}}\right)}}{1-\gamma^2}.
$$
\\
(see \cite{cessac:11a} for details)

\vspace{-10pt}

\ssu{Explicit calculation of the canonical Maximum Entropy Potential}\label{SExpansion}

The goal now is to derive from (\ref{PhiBMS}) a canonical potential $\H$  of the form (\ref{Hh}) whose spike interactions terms $h_l$'s are functions of the network parameters: the synaptic weight matrix $\cW$ and the external stimulus $\mathcal{I}$, $h_l \equiv h_l(\cW,\mathcal{I})$.

Equation (\ref{Livsic2}) gives a relation between the normalized potential and an equivalent non-normalized potential. From this equation, after considering the elimination of equivalent interactions is it possible to compute explicitly the values of the interaction terms $h_l$'s.\\
\\
\ssssu{Free energy:}\\
From (\ref{tpres}) and (\ref{PhiBMS}) we get the free energy: 
%
$$ - \phi (\om{0})=\mathcal{P}[\H]=-\sum_{k=1}^N \log\bra{1-\pi\pare{\frac{\theta-I_k \, \frac{1-\gamma^{D}}{1-\gamma}}{\sigma_B \sqrt{\frac{1-\gamma^{2D}}{1-\gamma^2}}}}}.$$
\\
\ssssu{Local fields:}\\
They are computed using equation (\ref{h_l}). We consider the periodic orbit obtained by the $R$-circular shift of the block corresponding to the monomial $\omega_i(D)$. We have therefore to compute $ \phi(\om{l_0}) + \phi (\om{l_1}) + \cdots + \phi (\om{l_D})$ using equation (\ref{PhiBMS}). To obtain this quantity we have compute $X_k$ (\ref{Xk}) for all the blocks in the periodic orbit. Note that $X_k$ does not depend on the last column of the blocks in the orbit. We abuse the notation by writing $\X{k}{\om{\sigma^nl}}$ instead of $\X{k}{\omega_0^{(\sigma^{n}l)D-1}}$. The same holds for $\et{kj}{\om{\sigma^n l}}$ and $\sigma_k(\om{\sigma^n l})$. We obtain:
\beq\label{Xkdeg1}
\scriptscriptstyle \X{k}{\om{\sigma^nl}} =
\left\{
\begin{array}{lllll}
\scriptscriptstyle \frac{\theta- W_{ki} \,\gamma^{n-1} -I_k\frac{1-\gamma^{D}}{1-\gamma}}{\sigma_B \sqrt{\frac{1-\gamma^{2D}}{1-\gamma^2}}}, \: 1 \leq n \leq R-1, \, k \neq i;\\
\scriptscriptstyle \frac{\theta- W_{kk} \,\gamma^{n-1} -I_k\frac{1-\gamma^{n}}{1-\gamma}}{\sigma_B \sqrt{\frac{1-\gamma^{2n}}{1-\gamma^2}}}, \: 1 \leq n \leq R-1, \, k = i;\\
\scriptscriptstyle \frac{\theta - I_k\frac{1-\gamma^D}{1-\gamma}}{\sigma_B \sqrt{\frac{1-\gamma^{2D}}{1-\gamma^2}}}, \, \forall k, \, n=0.
\end{array}
\right.
\eeq

Combining equations (\ref{h_l}) , (\ref{PhiBMS}) and (\ref{Xkdeg1}) we obtain:
\beq\label{h_IF_order1}
\begin{array}{lll}
\h_i =  
\sum_{n=1}^{R-1} \sum_{k =1}^N \log\bra{1 - \pi\pare{\X{k}{\om{\sigma^nl}}}} \, + \, \\
\qquad \sum_{k \neq i} \log\bra{1 - \pi\pare{\X{k}{\om{\sigma^0l}}}} \, + \, \\
\qquad \log\bra{\pi\pare{\X{i}{\om{\sigma^0l}}}} - R \phi(\omega^{(0)}).
\end{array}
\eeq
\\
which is an explicit function of synaptic weights and stimuli. Clearly:

\bit
\item The ``local field'' of a neuron $i$ depends non linearly on \textit{all} stimuli (not only $I_i$).

\item It depends non linearly on the incoming synaptic weights connected to $i$. 
\eit 

\ssssu{Pairwise interactions (instantaneous):}\\
We get:
\beq\label{Xkdeg2}
\scriptscriptstyle \X{k}{\om{\sigma^nl}} =
\left\{
\begin{array}{lllllll}
\scriptscriptstyle \frac{\theta- (W_{ki}+W_{kj}) \,\gamma^{n-1} -I_k\frac{1-\gamma^{D}}{1-\gamma}}{\sigma_B \sqrt{\frac{1-\gamma^{2D}}{1-\gamma^2}}}, \: 1 \leq n \leq R-1, \, k \neq i,j;\\
\scriptscriptstyle \frac{\theta- (W_{kk}+W_{kj}) \,\gamma^{n-1} -I_k\frac{1-\gamma^{n}}{1-\gamma}}{\sigma_B \sqrt{\frac{1-\gamma^{2n}}{1-\gamma^2}}}, \: 1 \leq n \leq R-1, \, k = i;\\
\scriptscriptstyle \frac{\theta- (W_{kk}+W_{ki}) \,\gamma^{n-1} -I_k\frac{1-\gamma^{n}}{1-\gamma}}{\sigma_B \sqrt{\frac{1-\gamma^{2n}}{1-\gamma^2}}}, \: 1 \leq n \leq R-1, \, k = j;\\
\scriptscriptstyle \frac{\theta - I_k\frac{1-\gamma^D}{1-\gamma}}{\sigma_B \sqrt{\frac{1-\gamma^{2D}}{1-\gamma^2}}}, \, \forall k, \, n=0.\\
\end{array}
\right.
\eeq

Plugging  (\ref{Xkdeg2}) in (\ref{PhiBMS}) and using (\ref{Jijising2}), one finally obtains $\J_{ij}$ as a explicit function of synaptic weights and stimulus. \\

Remarks:

\bit

\item The ``instantaneous pairwise" interaction $\J_{ij}$ depends not only on $W_{ij}$, but in all synaptic weights of neurons connected with $i$ or $j$.

\item It also depends in the stimulus of all neurons in the network.
 
\eit
\textbf{Pairwise interactions (1 time-step)}: \\
As mentioned in the previous section, in order to compute this term, the periodic orbit obtained by the $R$-circular shift of the block $\om{l_0}$ corresponding to the monomial $\omega_i(1)\omega_j(0)$ is not sufficient. We have therefore to use the periodic orbit obtained by our procedure \eqref{p.o}. From $\om{l_1}$ to $\om{l_4}$ we have already computed their corresponding value $ \X{k}{\om{\sigma^nl}}$ when computing the Local fields. From $\om{l_4}$ to $\om{l_{R+2}}$ we just circularly shift $\om{l_4}$. We compute the corresponding $ \X{k}{\om{\sigma^nl}}$:

\beq\label{Xkdeg21}
\scriptscriptstyle \X{k}{\om{\sigma^nl}} =
\left\{
\begin{array}{lllllll}
\scriptscriptstyle \frac{\theta- W_{ki}\gamma^{n-4}-W_{kj}\gamma^{n-3}
-I_k\frac{1-\gamma^{D}}{1-\gamma}}{\sigma_B
\sqrt{\frac{1-\gamma^{2D}}{1-\gamma^2}}}, \: 4 \leq n \leq R+2, \, k
\neq i,j;\\
\scriptscriptstyle \frac{\theta- W_{kk}\gamma^{n-4}-W_{kj}\gamma^{n-3}
-I_k\frac{1-\gamma^{n-3}}{1-\gamma}}{\sigma_B
\sqrt{\frac{1-\gamma^{2(n-3)}}{1-\gamma^2}}}, \: 4 \leq n \leq
R+2, \, k = i;\\
\scriptscriptstyle \frac{\theta- W_{kk}\gamma^{n-4}-W_{ki} \gamma^{n-3}
-I_k\frac{1-\gamma^{n-2}}{1-\gamma}}{\sigma_B
\sqrt{\frac{1-\gamma^{2(n-2)}}{1-\gamma^2}}}, \: 4 \leq n \leq R+2,
\, k = j;\\
\end{array}
\right.
\eeq
\\

We then apply equation (\ref{Livsic2}) to obtain the desired term, from previously computed interaction terms.\\

\begin{figure}[h]
\begin{center}
\scalebox{0.2}
{
\includegraphics{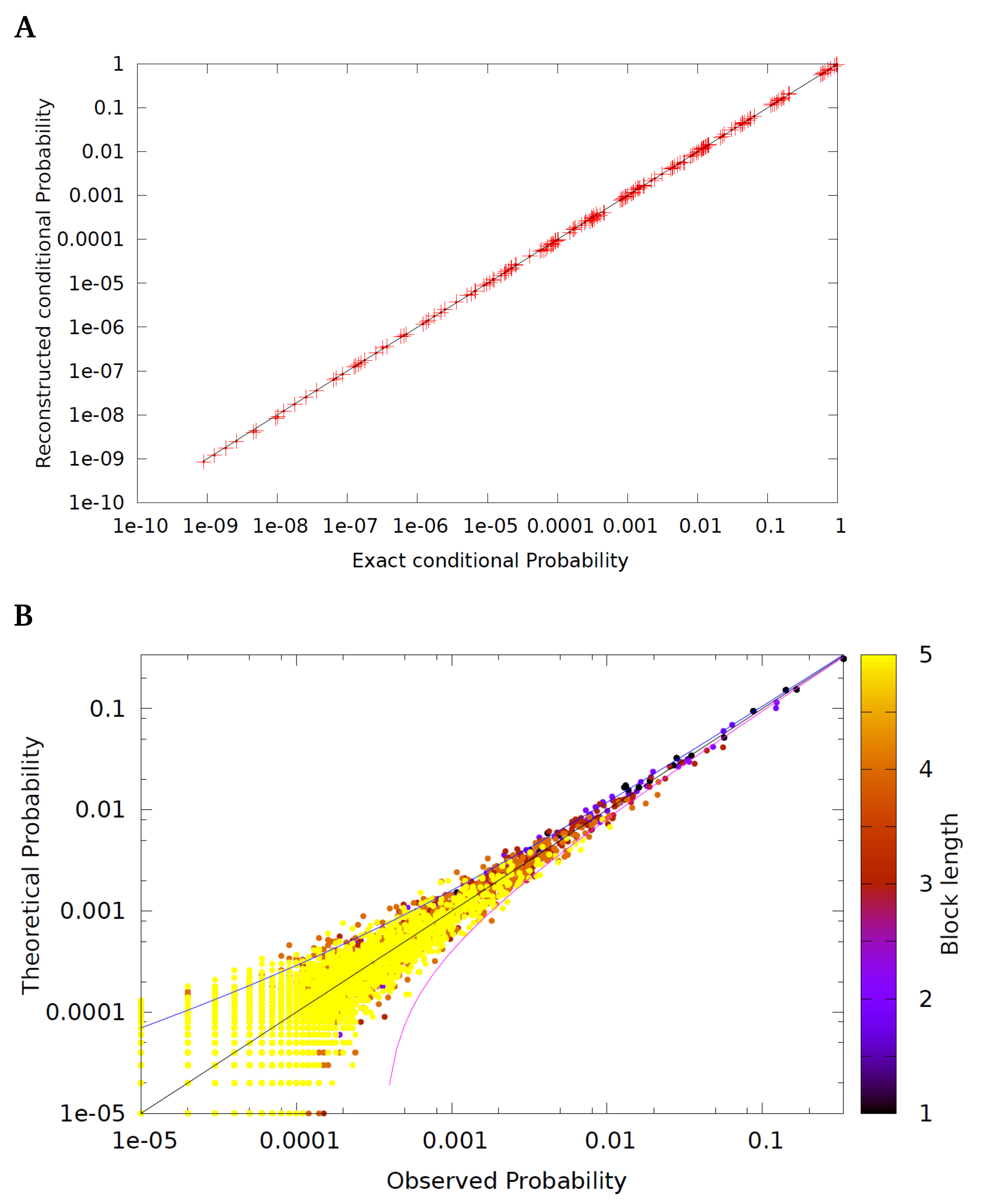}}
\caption{\label{fig:dual2} \footnotesize{(Color online) (A). Exact conditional probabilities for blocks of range $R$ obtained from the normalized potential (\ref{PhiBMS}), vs exact conditional probabilities associated with the potential (\ref{Hh}).  (B)  Empirical probabilities of blocks $\bloc{0}{k}$, $k=1, \dots 5$ (darker lower length), obtained from a discrete leaky integrate and fire spike train of size $T=10^5$ vs the probabilities of the same blocks predicted by the Gibbs ditribution with potential $\H$ (\ref{Hh}). Each dot stands for one of the $2^{Nk}$ spatio-temporal patterns, where $k$ is the block length. Diagonal shows equality. Confidence bounds (blue and red lines) correspond to fluctuations ruled by Central Limit Thorem. Plot is in log scale. This figure corresponds to $N=5,R=3, \gamma=0.2, \sigma_B=0.2, \theta=1, I_k=0.7 ,\ k=1, \dots 5$. The synaptic weights are random and sparse. Each neuron was randomly connected to other 2 neurons whose weights were drawn from a gaussian 0 mean and variance $\frac{J^2}{N}$. In this example $J=3$. \label{FHtestN5R3}}}
\end{center}
\end{figure}

A numerical illustration of our method is presented in figure (\ref{fig:dual2}). We start from the normalized potential (\ref{PhiBMS}) and construct the canonical equivalent potential. We then compare the conditional probability of patterns predicted by $\H$ with the empirical probabilities inferred from a spike train generated by (\ref{PhiBMS}). This is just an illustration, and not a systematic study. Note that this numerical analysis is limited to small $N,R$ since the number of terms in $\H$ grows exponentially fast, rendering intractable the method for $NR \geq 20$.

\section{V. ANSWERING QUESTION 3 AND GENERAL CONCLUSION}

When the normalized potential $\phi$ is derived from a neuro-mimetic model (e.g. eq. (\ref{PhiBMS})), it follows that the ``local fields"  $\h_i$ depends non linearly on the complete stimulus $\mathcal{I}$ (not only the stimulus applied to neuron $i$), and the synaptic weights matrix $\cW$. This is not that surprising. Even considering an Ising model of two neurons with no memory, a strong favorable pairwise
interaction between the two neurons will increase the average firing rate of both neurons, even in the absence of an external field. 
Likewise, $\J_{ij}$ depends on the whole  synaptic weights matrix $\cW$  and not only on the connection between $i$ and $j$. This example clearly shows that there is no straightforward relation
between the so-called ``functional connectivity" in Ising model $\J_{ij}$ and the neural synaptic connectivity ($W_{ij}$).\\

As stated in the introduction a neuro-mimetic model with $N$ neurons has $O(N^2)$ parameters, whereas a MaxEnt model with $N$ neurons and memory depth $D$ has $O(2^{NR})$ parameters $h_l$ (canonical potential). Since the correspondence from neuro-mimetic models to a MaxEnt potential is exact we have two possibilities:

\begin{enumerate}[(i)]
\item A large number of $h_l$'s vanish.
\item $h_l$'s are related among them. 
\end{enumerate}

The two possibilities are actually not exclusive. Let us first address this question from the mathematical (dynamical systems) viewpoint which was the line followed up to now.\\
Consider a neuro-mimetic model with a well defined dynamics (e.g. (\ref{Bms})) and the associated normalized potential $\phi=\phi(\mathcal{W},\mathcal{I})$ (e.g. (\ref{PhiBMS})). We may view a normalized potential as a point in a space: the coordinates of this point are fixed by $\mathcal{W},\mathcal{I}$. A neuro-mimetic model corresponds therefore to the space of normalized potential of dimension $O(N^2)$. Using the same representation MaxEnt models with memory depth $D$ span a space of dimension $O(2^{NR})$, but MaxEnt models equivalent to our neuro-mimetic model span a space of dimension $O(N^2)$. There is therefore a huge projection effect. Now eq (\ref{Jijising}), or, more generally, eq. (\ref{Livsic2}) show that (ii) always functionally holds: $h_l$ are non linear functions of $\mathcal{W},\mathcal{I}$ and are related to each others. This has a dramatic consequence. Assume that we want to fit (exactly) a neuro-mimetic model with a MaxEnt. We will need $2^{NR}$ terms whereas $O(N^2)$ are sufficient. This is exactly what happens in Fig 2. Now, we may have the hope that many $h_l$'s are zero or close to zero. This is actually where MaxEnt models could make a breakthrough, showing that, in real spike trains many $h_l$'s (almost) cancel would reveal a hidden law of nature. \\
What is the hope for this? If we address this question from the dynamical systems viewpoint, there is no hope. Indeed in this context, one has to look for generic conditions for $h_l$'s to vanish (case (i)). But it results from our analysis that the $h_l$'s of a canonical potential corresponding to a neuro-mimetic model are \textit{generically} non zero: considering e.g. \textit{random} synaptic weights $W_{ij}$, the probability that some $h_l$'s in (\ref{Livsic2}) vanishes is indeed zero. \\

However, real neural networks are non generic: synaptic weights are not drawn at random but result from a long phylogenetic and ontogenetic evolution. When trying to ``explain" spike statistics of real neural networks with the Maximum Entropy Principle, one is seeking some general laws that has to be expressed with relatively few phenomenological parameters in the potential (\ref{Hh}). The hope is that many coefficients coming from real data are $0$ or close to $0$. This could explain the efficiency of pairwise  MaxEnt models \cite{bialek-ranganathan:07} for spike trains analysis (although this effect could also arise due e.g binning). Our method  provides a way do detect this, if the l.h.s. in (\ref{Livsic2}) is close to $0$ \cite{cessac-cofre:13c}. 
%

Our method allows a mechanistic and causal understanding of the origin of correlations, in consequence, opens up new possibilities allowing a better understanding of the role of different neural network topologies and stimulus on the collective spike train statistics.



It is not limited to spike trains however and could also impact different areas of scientific knowledge where binary time series are considered. 

\section{ACKNOWLEDGEMENTS}
%
This work was supported by the French ministry of Research and University of Nice (EDSTIC), INRIA, ERC-NERVI number 227747, KEOPS ANR-CONICYT and European Union Project $\#$ FP7-269921 (BrainScales), Renvision $\#$ 600847. We thank the reviewers for constructive criticism.

\bibliographystyle{plain}
\bibliography{../odyssee,../biblio}

\end{document}